\documentclass[aps,pra,twocolumn,superscriptaddress,nofootinbib]{revtex4-2}
\usepackage{graphicx}
\usepackage{amssymb}
\usepackage{amsmath}
\usepackage{bm}
\usepackage{xcolor}
\usepackage{braket}
\usepackage{kotex}

\begin{document}

\title{Spin-Driven Stationary Turbulence in Spinor Bose-Einstein Condensates}

\author{Deokhwa Hong}
\affiliation{Department of Physics and Astronomy, and Institute of Applied Physics, Seoul National University, Seoul 08826, Korea}

\author{Junghoon Lee}
\affiliation{Department of Physics and Astronomy, and Institute of Applied Physics, Seoul National University, Seoul 08826, Korea}
\affiliation{Center for Correlated Electron Systems, Institute for Basic Science, Seoul 08826, Korea}

\author{Jongmin Kim}
\affiliation{Department of Physics and Astronomy, and Institute of Applied Physics, Seoul National University, Seoul 08826, Korea}

\author{Jong Heum Jung}
\affiliation{Department of Physics and Astronomy, and Institute of Applied Physics, Seoul National University, Seoul 08826, Korea}

\author{Kyuhwan Lee}
\affiliation{Department of Physics and Astronomy, and Institute of Applied Physics, Seoul National University, Seoul 08826, Korea}
\affiliation{Center for Correlated Electron Systems, Institute for Basic Science, Seoul 08826, Korea}

\author{Seji Kang}
\affiliation{Department of Physics and Astronomy, and Institute of Applied Physics, Seoul National University, Seoul 08826, Korea}

\author{Y. Shin}
\email{yishin@snu.ac.kr}
\affiliation{Department of Physics and Astronomy, and Institute of Applied Physics, Seoul National University, Seoul 08826, Korea}
\affiliation{Center for Correlated Electron Systems, Institute for Basic Science, Seoul 08826, Korea}
\affiliation{Institute of Applied Physics, Seoul National University, Seoul 08826, Republic of Korea}

\date{\today}

\begin{abstract}
We report the observation of stationary turbulence in antiferromagnetic spin-1 Bose-Einstein condensates driven by a radio-frequency magnetic field. The magnetic driving injects energy into the system by spin rotation and the energy is dissipated via dynamic instability, resulting in the emergence of an irregular spin texture in the condensate. Under continuous driving, the spinor condensate evolves into a nonequilibrium steady state with characteristic spin turbulence, while the low energy scale of spin excitations ensures that the sample's lifetime is minimally affected. When the driving strength is on par with the system's spin interaction energy and the quadratic Zeeman energy, remarkably, the stationary turbulent state exhibits spin-isotropic features in spin composition and spatial spin texture. We numerically show that ambient field fluctuations play a crucial role in sustaining the turbulent state within the system. These results open up new avenues for exploring quantum turbulence in spinor superfluid systems.
\end{abstract}
\maketitle

\section{Introduction}

Turbulence is a ubiquitous phenomenon in fluids, from classical to quantum; however, it is also a long-standing challenging problem because of its complexity. Atomic Bose-Einstein condensates (BECs), which enable exquisite control of experimental parameters and direct imaging of wave functions, provide a versatile platform for studying turbulence in quantum fluids~\cite{Paoletti11,White2014,Vengalattore11}. Moreover, the BEC system supports internal spin degrees of freedom, allowing for an extension of turbulence studies to spinor superfluids with multiple velocity fields. Owing to the rich symmetry of the order parameter manifold, the spinor BEC may host unconventional topological defects~\cite{Ueda12,Kurn13}, possibly leading to novel types of turbulence~\cite{Tsubota10,Tsubota14}. Many interesting phenomena were recently demonstrated with atomic BECs, such as energy cascades~\cite{Hadzibabic16, Hadzibabic22}, universal coarsening in decaying turbulence~\cite{Oberthaler15, Chin16}, giant vortex clusters~\cite{Neely19,Helmerson19}, as well as spin turbulence~\cite{Kurn06,Kim17,Kang17,Oberthaler18}.  

Studying turbulence often involves investigating a fluid that is continuously driven, resulting in a stationary turbulent state. Stationary turbulence refers to a state where the statistical properties of the turbulence remain the same over time, while the chaotic flow pattern changes spatiotemporally. Such stationary states facilitate theoretical descriptions of turbulence~\cite{Taylor35,Karman38, Ye_Zhou21}, making them an attractive setting for exploring complex far-from-equilibrium phenomena. In stationary turbulence, the inertial range for an energy cascade is fully developed, with a steady energy flow from forcing to dissipation~\cite{Kolmogorov41a,Kolmogorov41b,Obukhov41}. Therefore, it is highly desirable  to establish a proper experimental method of forcing that can generate stationary turbulent states.

In this paper, we report the observation of stationary turbulence in a spin-1 Bose-Einstein condensate under radio-frequency (RF) magnetic field driving. We continuously inject energy into the system via its spin channel, whose energy scale is one order of magnitude smaller than the characteristic energy scale of condensation, and observe that the system evolves into a long-lived, nonequilibrium steady state with turbulent flow. Furthermore, we find that under optimal driving conditions, the stationary turbulent system exhibits spin-isotropic features in spin composition and spatial texture. The observed nonequilibrium steady state represents a new class of superfluid turbulence, providing interesting opportunities for the study of quantum turbulence in a spinor superfluid.

\section{Spinor superfluid forced by spin rotation}

The quantum fluid studied in this work is a BEC of $^{23}$Na atoms in the $F$=1 hyperfine state with antiferromagnetic interactions~\cite{Zhou03,Seo15}. The mean-field spin energy of an unmagnetized $F$=1 BEC is given by 
\begin{equation}
\label{UndrivenEq}
    E_s = \Big[\frac{c_2 n}{2} \langle \mathbf{F} \rangle^2 + q \langle F_z^2 \rangle \Big] n,
\end{equation} 
where $c_2$ is the spin-dependent interaction coefficient, $n$ is the atom number density, $\mathbf{F} = (F_x,F_y,F_z)$ is the spin operator of the spin-1 system, and $q$ is the quadratic Zeeman energy. For $^{23}$Na, $c_2>0$~\cite{Ketterle98} and the mean-field ground state is a polar state with $\langle \hat{\bm{d}}\cdot \mathbf{F}\rangle=0$, where $\hat{\bm{d}}$ is a unit vector, called the director, indicating the quantization axis along which the system is in the $m_F$=0 state~\cite{Ueda12}. The quadratic Zeeman energy imposes spin anisotropy, giving $E_s=q (1-(\hat{\bm{d}}\cdot\hat{\bm{z}})^2)$, where $\hat{\bm{z}}$ denotes the direction of the external magnetic field. For $q>0$, the ground state is the easy-axis polar (EAP) state with $\hat{\bm{d}}\parallel \hat{\bm{z}}$~\cite{Raman11}.

\begin{figure}[t]
	\includegraphics[width=8.4cm]{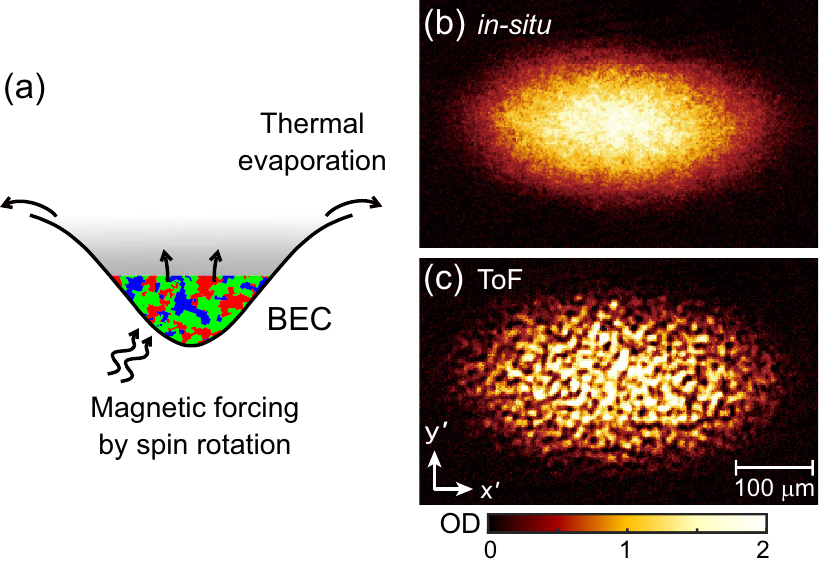}
	\caption{Stationary turbulence in a spinor Bose-Einstein condensate (BEC) under magnetic forcing. (a) Schematic illustration of the energy flow in a driven BEC system. Turbulence is sustained by driving with a radio-frequency (RF) magnetic field and energy dissipation into a thermal gas whose temperature is regulated by evaporation cooling due to the finite trap depth. The color pattern of the BEC indicates its irregular spin texture. Absorption images of turbulent BEC after 5-s driving with the RF field, taken (b) {\it in-situ} and (c) after a 18-ms time-of-flight (ToF). The Rabi frequency of the RF driving was $\Omega/2\pi=150$~Hz.}
	\label{Fig1}
\end{figure}

When the system is excited with a spin texture that differs from the ground state, it relaxes back to the ground state. Recent experiments have shown that during this relaxation process, spin turbulence emerges due to the dynamic and energetic instabilities of the excited state~\cite{Kang17,Kang20,Raman11}. This involves spontaneous generation of spin waves via spin exchange collisions and phase separation between spin components~\cite{Ketterle98, Gerbier19}, resulting in complex velocity fields of the spin components. Eventually, the turbulence subsides as the system returns to its ground state~\cite{Kang17,Kang20}. However, if energy could be continuously injected in a way to drive the spin texture out of the ground state, a nonequilibrium steady turbulent state might emerge from the balance between spin driving and relaxation [Fig~\ref{Fig1}(a)]. In this work, we demonstrate that such continuous forcing can be achieved by slowly rotating the spins using an external RF magnetic field in conjunction with small background field fluctuations.

\begin{figure*}[t]
	\includegraphics[width=16.7cm]{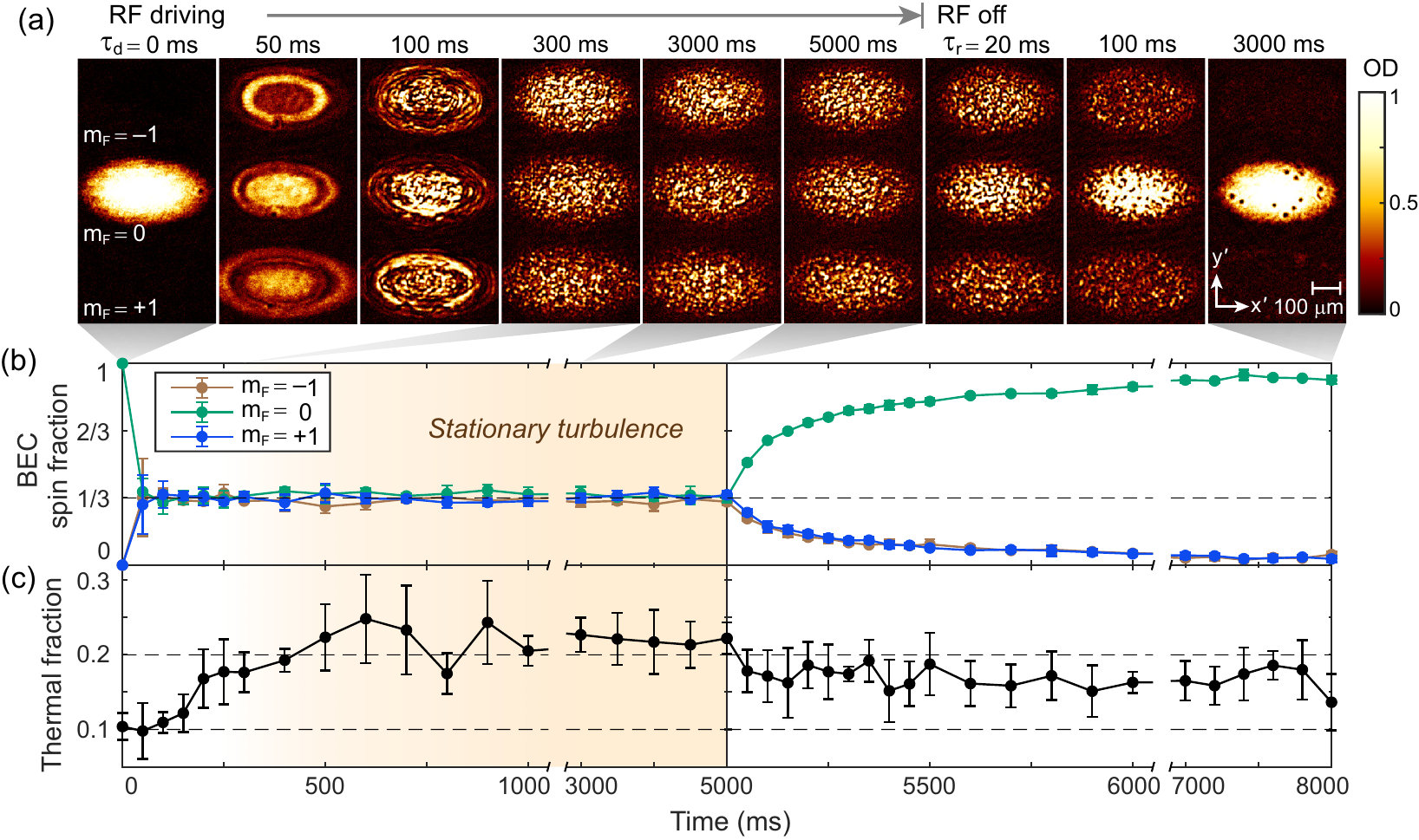}
	\caption{Turbulence generation and relaxation in a $F$=1 spinor BEC. (a) ToF images of samples with Stern-Gerlach (SG) spin separation for various RF driving times, $\tau_d$, with  $\Omega/2\pi=150$~Hz and for various relaxation times, $\tau_r$, after 5-s driving. The BEC is initially prepared in the $m_F$=0 state. An irregular spin texture develops within about 300 ms and the turbulent state becomes stationary thereafter. With the RF driving turned off, the BEC relaxes into the $m_F$=0 ground state, containing quantum vortices as topological defects. (b) Time evolution of the fractional spin populations in the BEC and (c) thermal fraction of the sample. Data points indicate mean values of five to ten measurements and error bars indicate their standard deviations.}
	\label{ByTime}
\end{figure*}

\section{Results and dicussion}

\subsection{Experiement setup}

We start our experiment with preparing a BEC of about 8$\times$10$^{6}$ $^{23}$Na atoms in the $|F$$=$$1, m_{F}$$=$$0\rangle$ state in an optical dipole trap. The condensate has highly oblate geometry and its Thomas-Fermi (TF) radii are $(R_{x'},R_{y'},R_{z'})$~$\approx$~$(230, 110, 2.4)~\mu$m for the trapping frequencies of $(\omega_{x'},\omega_{y'},\omega_{z'}) \approx 2\pi\times(4.3, 8.8, 420)$~Hz, where $x', y'$ and $z'$ denote spatial coordinates. For the peak atomic density $n_0$, the spin interaction energy is $c_2n_0 \approx h \times 39$~Hz and the spin healing length is $\xi_s\approx2.4~\mu$m~\cite{Tiemann11}, which is comparable to the sample thickness. A uniform external magnetic field of $B_0\approx 0.41$~G is applied along $\hat{\bm{z}}=(-\hat{\bm{x}}'+\hat{\bm{y}}')/\sqrt{2}$ and the quadratic Zeeman energy is $q=\alpha B_0^2 \approx h\times 47$~Hz with $\alpha=h \times 277  \text{Hz}/\text{G}^2$ for $^{23}$Na in the $F$=1 state.  The field gradient in the sample plane is controlled to be less than 0.1~mG/cm. 

%% Experimental method
To rotate the spin state, we apply an RF magnetic field oscillating along $\hat{\bm{y}}'$. The oscillating frequency $\omega$ is set to be at the Larmor frequency  $\omega_0=g_F \mu_\text{B} B_0/\hbar \approx 2\pi\times 291$~kHz, where $\mu_\text{B}$ is the Bohr magneton and $g_F=\frac{1}{2} $ is the Land\'{e} $g$-factor of the atom, and the RF field drives the atoms in the $m_F$=0 state to the $m_F$=$\pm1$ state. In the experiment, the Rabi frequency, $\Omega/2\pi$, of the RF driving ranges from 3~Hz to 15~kHz. It is important to note that ambient magnetic field fluctuations, $\delta \mathbf{B}(t)$, may affect the RF driving. In a rotating frame, taking the rotating wave approximation, the single-particle spin Hamiltonian is expressed as 
\begin{equation}
 H_s = -\hbar \delta(t) F_z + q F_z^2 -\hbar \Omega F_x,
 \end{equation}
 where $\delta(t)=g_F \mu_{\text{B}}\delta B_z/\hbar$\footnote{Transverse field fluctuations are ignored in the rotating wave approximation.}. This suggests that the effect of field fluctuations can be interpreted as a random wobbling of the spin rotation axis. In our experiment, we estimated the magnitude of the magnetic field fluctuations to be approximately 1~mG~(see Appendix~\ref{sec:Appendix1}), corresponding to $\delta \approx 2\pi\times 0.7$~kHz. This value is non-negligible considering our range of $\Omega$.

\subsection{Long-lived stationary turbulence}

We detect turbulence in the BEC by taking an absorption image of the sample after a time-of-flight (ToF) expansion. During a ToF, the internal turbulent flow leads to development of irregular density modulations in the freely expanding condensate~\cite{Choi12}. Figure~\ref{Fig1}(b) shows an absorption image of the sample after 5-s RF driving with $\Omega/2\pi=150$~Hz and an 18-ms ToF. A clearly visible, irregular density pattern appears, indicating turbulence generation in the driven BEC. We note that the {\it in-situ} density profile of the driven BEC was smooth, confirmed by imaging without the ToF.

We investigate the time evolution of the BEC upon application of magnetic driving by taking ToF images with Stern-Gerlach (SG) spin separation for various driving times [Fig.~\ref{ByTime}(a)]\footnote{Due to the inhomogeniety of the field gradient for SG spin separation, the clouds of $m_F$=$\pm1$ spin components are slightly extended or squeezed differently.}. In the early stage of evolution, the BEC shows rather long wavelength excitations with the same geometry as the sample~\cite{Klempt10} and then, the excitations break into smaller segments, revealing a cascade characteristic of turbulence~\cite{Kang17}. A completely irregular spin texture develops within 300~ms over the whole condensate and is maintained thereafter. In the steady state, the fractional populations, $\eta_{0,\pm1}$, of the spin states are equalized at $1/3$ [Fig.~\ref{ByTime}(b)], where $\eta_{s}=N_s/N_c$ with $N_s$ being the number of condensed atoms in the $m_F$=$s$ state ($s\in\{0,\pm 1\}$) and $N_c=\sum_{s} N_s$. As the turbulence is generated, the thermal fraction, $\eta_{th}$, of the sample increases accordingly and saturates after about 500~ms [Fig.~\ref{ByTime}(c)]. Here, $\eta_{th}=N_{th}/N_t$ with $N_{th}$ being the number of thermal atoms and $N_t$=$N_{th}$+$N_c$\footnote{The thermal fraction was measured by analyzing the ToF images without SG, from a fit of the outer wings of the sample to a Gaussian function.}. The increase in $\eta_{th}$ indicates that the injected energy eventually dissipates into the thermal gas coexisting with the BEC, whose temperature is regulated by evaporation cooling due to the finite trap depth [Fig.~\ref{Fig1}(a)].

When the RF driving is turned off after its 5-s application, the turbulent BEC relaxes into the EAP ground state, as expected, in which $\eta_0$ monotonically increases toward unity and the spin texture is coarsened, leaving quantum vortices as topological defects [Fig.~\ref{ByTime}(a)]. Meanwhile, the thermal fraction gradually decreases to a new equilibrium value after stopping the energy injection [Fig.~\ref{ByTime}(c)].  

\begin{figure} [b]
	\includegraphics[width=8.6cm]{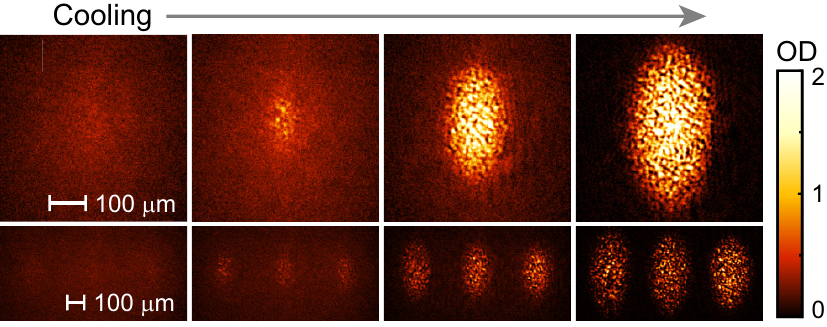}
	\caption{
	Direct formation of a spin-turbulent BEC under magnetic driving. A thermal gas is evaporatively cooled in the presence of RF field with $\Omega/2\pi$=150~Hz and a BEC is formed and grows at low temperatures. (upper row) ToF images of samples for various temperatures and (bottom row) the corresponding images with SG spin separation.}
	\label{BECforming}
\end{figure}

%\subsection{ Low heating of spin driving}

The turbulent BEC exhibits longevity. In our experiment, the $1/e$ lifetime of the BEC is reduced from 25.1~s to 22.0~s, only by about 10\%, when subject to RF driving at $\Omega/2\pi$=150~Hz. We attribute this low heating by RF driving to the low energy scale of spin excitations, which is about 30 times smaller than that of the density excitations representing the system's condensation energy. In the relaxation process shown in Fig.~\ref{ByTime}(b), the initial increase rate of $\eta_0$ is about 2.7~s$^{-1}$, which suggests an energy dissipation rate of $q \frac{d\eta_0}{dt}$$\sim$$ k_\textrm{B}$$\times 6$~nK/s per atom. However, from the condensate chemical potential $\mu\approx k_\textrm{B} \times 50$~nK and the measured minor reduction in BEC lifetime, we expect significantly lower energy dissipation under RF driving conditions. Notably, we observe that even in the presence of RF driving, a thermal gas can be effectively cooled through forced evaporation to directly produce a turbulent BEC~(Fig.~\ref{BECforming}).

\subsection{Driving strength dependence}

\begin{figure}[t]
	\includegraphics[width=8.6cm]{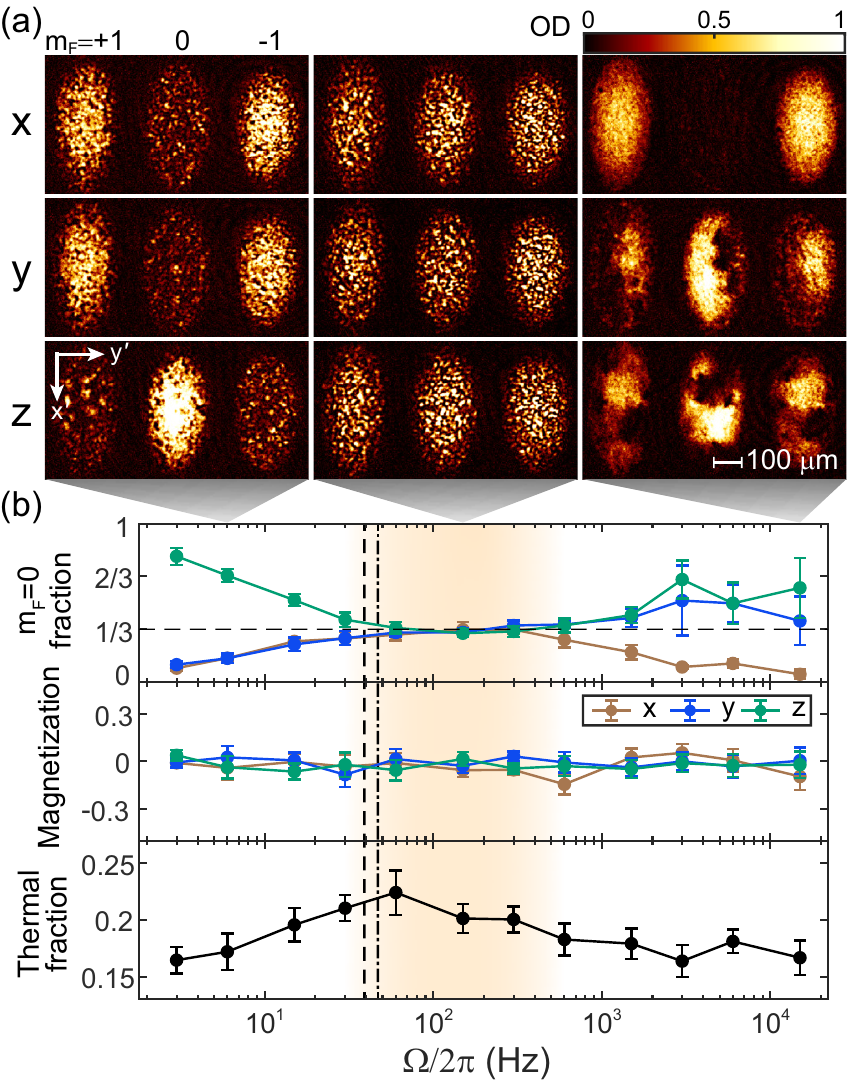}
	\caption{Evolution of the stationary turbulent state for various RF driving strengths. (a) SG images for $x$, $y$, and $z$ spin axes after 5-s driving with $\Omega/2\pi$ = 6 Hz, 150 Hz, and 15 kHz. The $z$ axis denotes the direction of the external magnetic field and the $x$ axis denotes the spin rotation axis for the driving RF field. (b) $m_F$=0 spin fractions $\eta_{0,\alpha}$ and normalized magnetizations $M_\alpha$ of the turbulent BEC for $\alpha=x,y,z$ spin axes, and the thermal fraction $\eta_{th}$ of the sample as functions of $\Omega$. Data points were obtained by averaging over seven to fourteen measurements with 5-s driving and error bars indicate the standard deviations of the measurements. The vertical dash line and dash-dot line indicate the spin interaction energy $c_2 n$ and the quadratic Zeeman energy $q$, respectively. The shaded area indicates the driving range for spin-isotropic turbulence with $\eta_{0x}=\eta_{0y}=\eta_{0z}=\frac{1}{3}$.}
	\label{ByOmega}
\end{figure}

As stationary turbulence is sustained by a dynamical balance between RF driving and the relaxation of the system, the resulting turbulence changes as the driving power varies. We characterize the turbulent state and its dependence on the driving power by taking SG images of the sample for various $\Omega$, along the three orthogonal spin axes, $x$, $y$ and $z$ [Fig.~\ref{ByOmega}(a)]. Here, the spin rotation axis defined by the oscillating RF field is denoted as $\hat{\bm{x}}$. For $x$ and $y$ axis imaging, an additional, short RF pulse of 60~$\mu$s is applied right before releasing the sample from the trap to rotate the corresponding spin axis to $\hat{\bm{z}}$, the imaging axis direction. From the SG images for each spin axis ($\alpha\in \{x,y,z\}$), we determine the fractional populations $\eta_{0,\alpha}$ and the normalized net magnetization $M_{\alpha}$=$\eta_{1,\alpha}-\eta_{-1,\alpha}$ of the condensate. Note that magnetization is not a conserved quantity in this RF driven system because $[H_s, \hat{\bm{u}}\cdot \mathbf{F}] \neq 0$ for all unit vectors $\hat{\bm{u}}$.

Figure~\ref{ByOmega}(b) displays the measurement results of $\eta_{0,\alpha}$ and $M_\alpha$ for 5-s driving as functions of $\Omega$, together with the thermal fraction $\eta_{th}$. As $\Omega$ increases, the spin composition $\{\eta_{0x},\eta_{0y},\eta_{0z}\}$ gradually departs from that of the EAP ground state, $\{0,0,1\}$, while keeping $\eta_{0x} = \eta_{0y}$. When $\Omega>\Omega_{c1}\approx2\pi\times 50$~Hz, it converges to $\{\frac{1}{3},\frac{1}{3},\frac{1}{3}\}$, which is a maximally mixed state. It is noticeable that the threshold frequency $\Omega_{c1}$ is close to $q$ and $c_2 n$, which determine the spin energy scale of the system. During its evolution, net magnetization is not developed in the system, keeping $M_\alpha\approx 0$.

%% Strong driving
When the RF driving is further strengthened with $\Omega>\Omega_{c2}\approx 2\pi \times 500$~Hz, we observe that $\eta_{0x}$ decreases and eventually vanishes. In the $x$-axis image, the $m_F=\pm 1$ components exhibit nearly identical density distributions, while in the $y$ and $z$-axis images, large spin domain structures are observed  [Fig.~\ref{ByOmega}(a) right]. These observations indicate that the BEC evolves into into an easy-plane polar (EPP) state with $\hat{\bm{d}}\perp \hat{\bm{x}}$. The spin fractions $\eta_{0y}$ and $\eta_{0z}$ show mean values of approximately 1/2 with large variances, consistent with the EPP state having the rotation symmetry along the $x$ axis. This change can be attributed to the time-averaging effect induced by fast spin rotation. In the regime where $\Omega\gg q/\hbar$, the atomic spins rotate rapidly within the $yz$ plane, resulting in a time-averaged effective spin energy of a polar state as $E_{s,\textrm{eff}}=\frac{q}{2}(1+(\hat{\bm{d}}\cdot\hat{\bm{x}})^2)$. Consequently, the director $\hat{\bm{d}}$ becomes effectively confined in the $yz$ plane.

The thermal fraction of the system reaches its peak at $\Omega \approx \Omega_{c1}$ and gradually decreases as $\Omega$ increases beyond $\Omega_{c1}$. This behavior is in agreement with the observed evolution of the turbulence magnitude in the system.

\subsection{Role of field fluctuations}

The sustainability of the turbulent state under continuous RF field raises intriguing questions regarding the new ground state under the driving and  how the system maintains its non-equilibrium state. In Appendix~\ref{sec:Appendix2}, we provide a phase diagram of the mean-field spin ground state for a homogeneous BEC driven by RF field. The diagram reveals a continuous transition from the EAP state with $\hat{\bm{d}}\parallel \hat{\bm{z}}$ to a ferromagnetic state with spin polarization along $\hat{\bm{x}}$ as the driving strength increases. This transition is governed by the competition between the quadratic Zeeman energy $q$ and the linear Zeeman energy $\hbar \Omega$. Based on this mean-field phase diagram, it is expected that the system would relax to its ground state through a transient period of turbulence following the sudden application of RF driving.

To gain insights into the sustaining mechanism of turbulence, we numerically investigate the dynamics of the spin-1 BEC under RF driving, taking into account realistic experimental conditions such as density inhomogeniety \cite{Ferrari21} and magnetic field fluctuations, which might facilitate the spin disordering of the BEC in spatial and temporal manners, respectively. Our numerical study is based on the two-dimensional Gross-Pitaevskii equation (GPE) for the spinor BEC,
\begin{equation}
    i\hbar \partial_{t} \Psi = \Big[ -\frac{\hbar^2}{2m}\nabla^2 +  H_s+V_{\text{trap}}(\textbf{r}')+c_0n+c_2n\langle \textbf{F} \rangle \cdot\textbf{F}-\mu \Big] \Psi,
\end{equation}
where $\Psi = (\psi_{+1}, \psi_0, \psi_{-1})^\textrm{T}$ with $\psi_s$ being the wave function of the $m_F$=$s$ component, $m$ is the atomic mass, $V_{\text{trap}}(\textbf{r}') = \frac{1}{2}m(\omega_{x'}^2x'^2+\omega_{y'}^2 y'^2)$ is the harmonic trapping potential, and $c_0$ is the spin-independent interaction coefficient.
The system parameters are chosen to be close to experimental values; $q/h$ = 47~Hz, $\mu/h$=1242~Hz ($c_2n_0/h$ = 45~Hz), $c_0/c_2=27.6$, $(R_{x'}, R_{y'})= (113,58)\xi_s$. The initial state of the BEC is a stationary EAP state with small Gaussian noises and the GPE is solved using a relaxation pseudo spectral scheme~\cite{Besse04, Antoine15}.

\begin{figure} [t]
	\includegraphics[width=8.6cm]{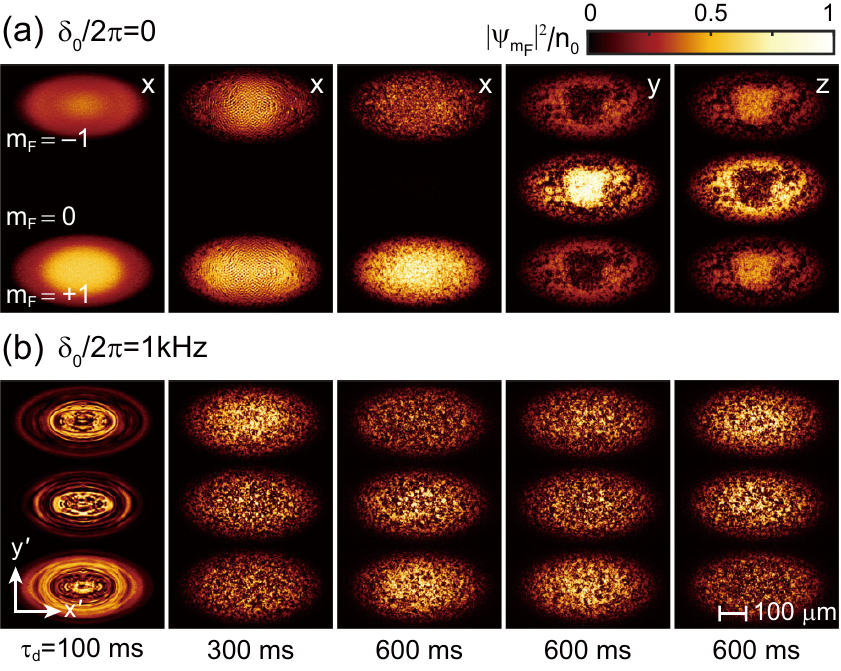}
	\caption{Effect of magnetic field fluctuations. Numerical simulation results of the density distributions of the spin components under RF driving for $\Omega/2\pi=150$~Hz (a) without and (b) with magnetic field fluctuations. Field fluctuations were modeled to be $\delta(t)=\delta_0 \sin (2\pi f t)$ with $\delta_0/2\pi=1$~kHz and $f=60$~Hz. $x$, $y$, and $z$ denote the spin projection axes.}
	\label{NumericEvolution}
\end{figure}

In Fig.~\ref{NumericEvolution}(a), we show the time evolution of the BEC without field noises ($\delta(t)$=0) for $\Omega/2\pi=150$~Hz. As spin excitations develop, magnetization along $\hat{\bm{x}}$ increases, consistent with the mean-field phase diagram. The spin fraction $\eta_{0x}$ remains zero throughout the evolution because the $m_x$=0 state is an eigenstate of the driven system. However, when field fluctuations are introduced, the dynamics of the BEC are qualitatively altered [Fig.~\ref{NumericEvolution}(b)]. In this case, we model the field fluctuations as a sinusoidal function $\delta (t) = \delta_{0}\sin(2\pi f t)$ with $\delta_{0}/2\pi = 1$~kHz and $f = 60$~Hz, reflecting the experimental conditions. An irregular spin texture involving all three spin components emerges, resembling the experimental observations. These numerical results clearly indicate the significant role of field fluctuations in generating the stationary, isotropic turbulence state. 

The interplay of RF driving and field fluctuations in turbulent generation may be speculated as follows: the RF driving brings the system to a dynamically critical state that is extremely sensitive to external magnetic field variations~\cite{Rautenberg20}, or field fluctuations render the system's ground state temporally ill-defined to make the relaxation process last effectively forever. It is noteworthy that the upper bound $\Omega_{c2}$ for the region of stationary turbulence region in Fig.~\ref{ByOmega}(b) is comparable to the estimated magnitude of field noise in terms of the Larmor frequency. This might imply that for high $\Omega>\Omega_{c2}$ the system becomes robust against field noises. The specific mechanisms by which field fluctuations contribute to turbulence generation warrants further investigation in future studies.

\begin{figure}[t]
	\includegraphics[width=8.4cm]{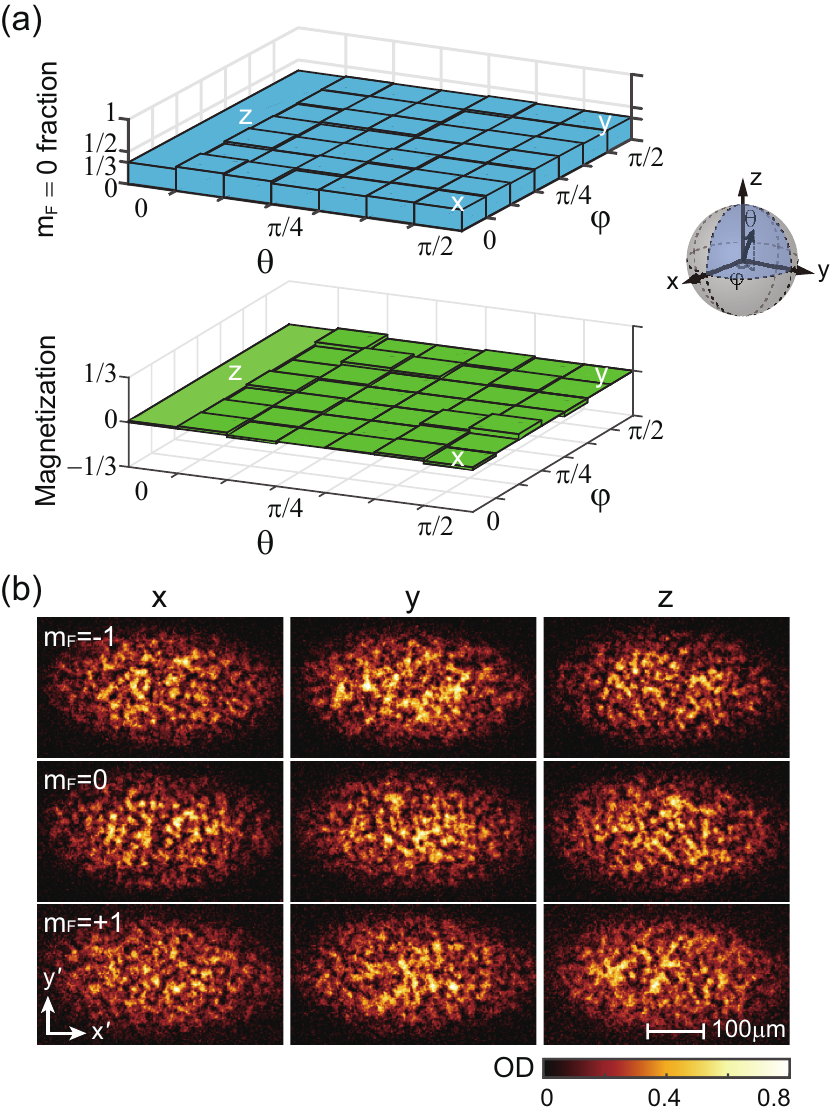}
	\caption{Spin isotropy of the turbulence. (a) $m_F$=0 spin fraction and magnetization for various quantization axes. $\theta$ and $\phi$ denotes the polar and azimuthal angles of the measurement axis. (b) {\it in-situ} images of the $m_F$=$-1,0,+1$ spin components for $x$, $y$, $z$ axes, after 5-s driving with $\Omega/2\pi$=150~Hz and $q/h=47$~Hz. In the spin-selective {\it in-situ} imaging, a fraction of atoms in the target spin state was rapidly transferred to the $F$=2 state with a short microwave pulse of 120~$\mu$s and imaged by a $F$=2 resonant light.
    }
	\label{Spinisotropy}
\end{figure}

\begin{figure}
	\includegraphics[width=7.9cm]{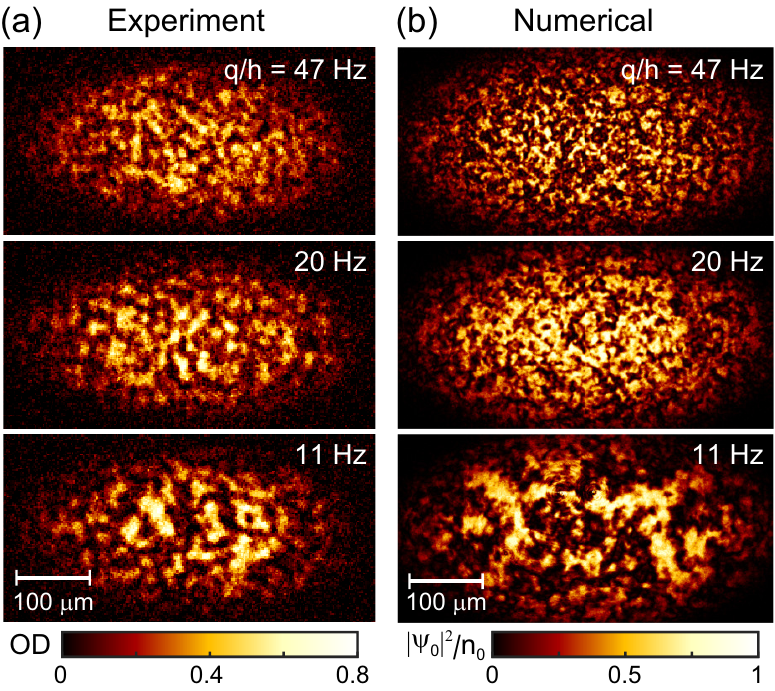}
	\caption{Turbulent BECs for various quadratic Zeeman energies.  (a) {\it in-situ} $z$-axis images of the $m_F$=0 spin component after 5-s driving with $\Omega/2\pi$=150~Hz for $q/h=47$~Hz, 20~Hz, and 11~Hz.  (a) Numerical simulation results for the same $q$ values  after 300-ms magnetic driving with $\Omega/2\pi$=150~Hz and $\delta_0/2\pi$=1~kHz.}
	\label{Numeric_q}
\end{figure}

\subsection{Spin-isotropic turbulence}

The main finding of this study is the emergence of a spin-isotropic, stationary turbulent state within the intermediate driving regime of $\Omega_{c1}<\Omega<\Omega_{c2}$. The spin-isotropic nature refers to the equal spin fractions observed along all spin axes and it indicates a maximum entanglement entropy of the spinor BEC when viewing the system as a composite of spin and spatial dimensions. To further validate the spin-isotropic property of the turbulence, we extend our measurements of the fractional populations of the spin states to various quantization axes. The control of quantization axis is achieved by adjusting the pulse durations of short RF pulses applied right before releasing the sample. We observe that $\eta_{0,\pm1}\approx \frac{1}{3}$ in all directions [Fig.~\ref{Spinisotropy}(a)], providing strong evidence for the spin-isotropic turbulence in the spinor BEC. Importantly, we find that such a turbulent state can be generated regardless of the initial spin state of the BEC.

Furthermore, we investgate the spin texture of the turbulent BEC by performing \emph{in-situ} images of the $m_F$=0 component along all spin axes [Fig.~\ref{Spinisotropy}(b)].  In the imaging, a fraction of atoms in the $|F$=1,$m_F$=0$\rangle$ state was rapidly transferred to the $|F$=2, $m_F$=0$\rangle$ state using a short microwave pulse of 120~$\mu$s and then imaged by a $F$=2 resonant light. While the quantitative analysis of the density distributions is limited by imaging noise, the spatial structures of the spin domains appear similar for different imaging axes, suggesting the isotropic character of the spin texture of the turbulent state.

In addition, we make an interesting observation that the spatial domain size increases as $q$ decreases as shown in Fig.~\ref{Spinisotropy}(c). This observation highlights the role of the intrinsic spin-anisotropy of the system in spin texture formation. The similar $q$ dependence of the characteristic length scale of spin turbulence has been observed in previous quench experiments~\cite{Kang17,Kang20}. To further support our experimental observations, we conduct numerical simulations for various $q$ values and find good agreement between the simulation results and experimental data [Fig.~\ref{Numeric_q}(a)].

\section{Conclusion}

We have observed the generation of stationary turbulence in a magnetically driven spin-1 BEC system, which has intrinsic instability due to spin anisotropy. Under optimal driving conditions, spin-isotropic turbulent states emerge, exhibiting the same spin compositions for all quantization axes. Thanks to its long lifetime, we expect that the observed stationary turbulence state will provide interesting opportunities for studies of nonequilibrium spinor superfluid dynamics. An immediate extension of this work would be to quantitatively characterize the spin texture~\cite{Tsubota14} and energy flow~\cite{Tsubota06,Knolle21} of the turbulent state and to investigate the nature of the superfluidity of this spin-disordered, turbulent BEC~\cite{Moore06,Evrard21}.

\begin{acknowledgments}
We thank Sang Won Seo and Joon Hyun Kim for their early contributions to the construction of the experimental apparatus, and Sol Kim for experimental assistance. This work was supported by the National Research
Foundation of Korea (NRF-2018R1A2B3003373, NRF-2019M3E4A1080400, NRF-2019H1A2A1074494) and the Institute for Basic Science in Korea (IBS-R009-D1).

\end{acknowledgments}

\appendix

\section{Field noise estimation}\label{sec:Appendix1}

\begin{figure}[t]
	\includegraphics[width=8.0cm]{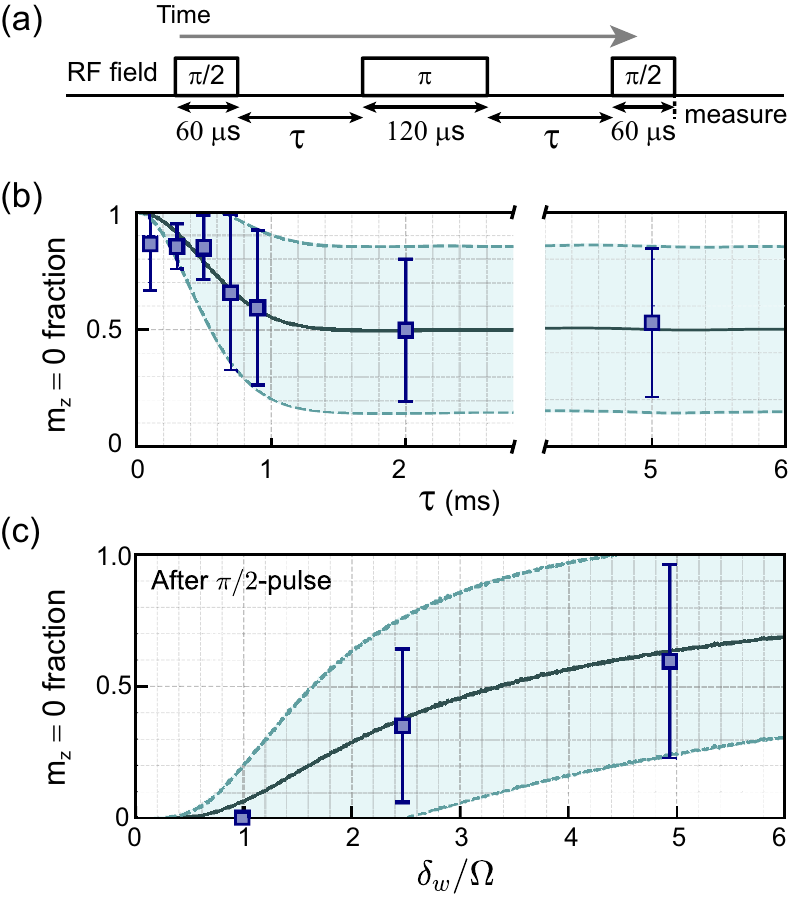}
\caption{Estimation of magnetic field noises. (a) RF pulse sequence for the spin echo measurement of magnetic field noises. The sample was initially prepared in the $m_z$=0 state and its spin fraction was measured right after the third RF pulse. (b) $m_z$=0 spin fraction as a function of the interrogation time $\tau$. Squares are the mean value of 20 measurements and the error bars are their standard deviation. (c) $m_z$=0 fraction after a $\pi/2$ RF pulse as a function of $\delta_w/\Omega$ ($\delta_w=2\pi\times 740$~Hz). Squares indicate the mean value of 5 experimental measurements and the error bars denote their standard deviation. The spin dynamics in the measurements was numerically simulated for Gaussian random noises. In (b) and (c), solid lines and shaded regions indicate the mean value and standard deviation of about $10^4$ numerical simulation results for the field noises with width of $\delta_w=2\pi\times 740$~Hz.}
	\label{ACnoise}
\end{figure}

To estimate the magnitude of the magnetic field fluctuations in the experiment, we measured the spin dephasing time using a spin echo scheme. The RF pulse sequence is described in Fig~\ref{ACnoise}(a). A BEC sample was initially prepared in the $m_z$=0 state and three RF pulses with $\omega$=$\omega_0$ were sequentially applied with the same time intervals of $\tau$, which realized spin rotations of $\pi/2$, $\pi$, and $\pi/2$ along $\hat{\bm{x}}$, respectively. By the first pulse, the atomic spin state was transferred to a superposition of the $m_z$=1 and $m_z$=$-1$ states, and then, the magnetic field fluctuations affect the evolution of the relative phase between the two Zeeman states. By the second pulse, the spin state was flipped along $\hat{\bm{x}}$ and the dephasing effect during the first inter-pulse period due to the spatial inhomogeneity of the external magnetic field was reversed during the second inter-pulse period. Right after the third RF pulse, the fractional spin population $\eta_{0z}$ of the $m_z$=0 state was measured, whose variations result from the temporal magnetic filed fluctuations over the whole sequence time.

%\begin{figure}
%	\includegraphics[width=8.6cm]{Fig9n}
%	\caption{$m_z$=0 fraction after a $\pi/2$ RF pulse as a function of $\delta_w/\Omega$ ($\delta_w=2\pi\times 740$~Hz). Squares indicate the mean value of 5 experimental measurements and the error bars denote their standard deviation. Solid line and shaded region indicate the mean value and standard deviation of $5\times10^4$ numerical simulation results for Gaussian field noises.}
%	\label{DCnoise}
%\end{figure}

The measurement results are shown in Fig.~\ref{ACnoise}(b) as a function of $\tau$. As $\tau$ increases, the mean value of $\eta_{0z}$ rapidly decreases from 1 to 0.5 and its variations increase accordingly. The standard deviation of the measured values reaches the value of $\frac{1}{2\sqrt{2}}$ for fully dephased states, within a few ms.
To determine the magnitude of the field fluctuations, we numerically simulate the spin dynamics by modeling $\delta(t)$ with Gaussian random noises. By fitting the simulation results to the experimental data, we obtain a best Gaussian width of $\delta_w=2\pi\times 740$~Hz, which corresponds to field fluctuations of about 1~mG.

\begin{figure}[t]
    \includegraphics[width=7.7cm]{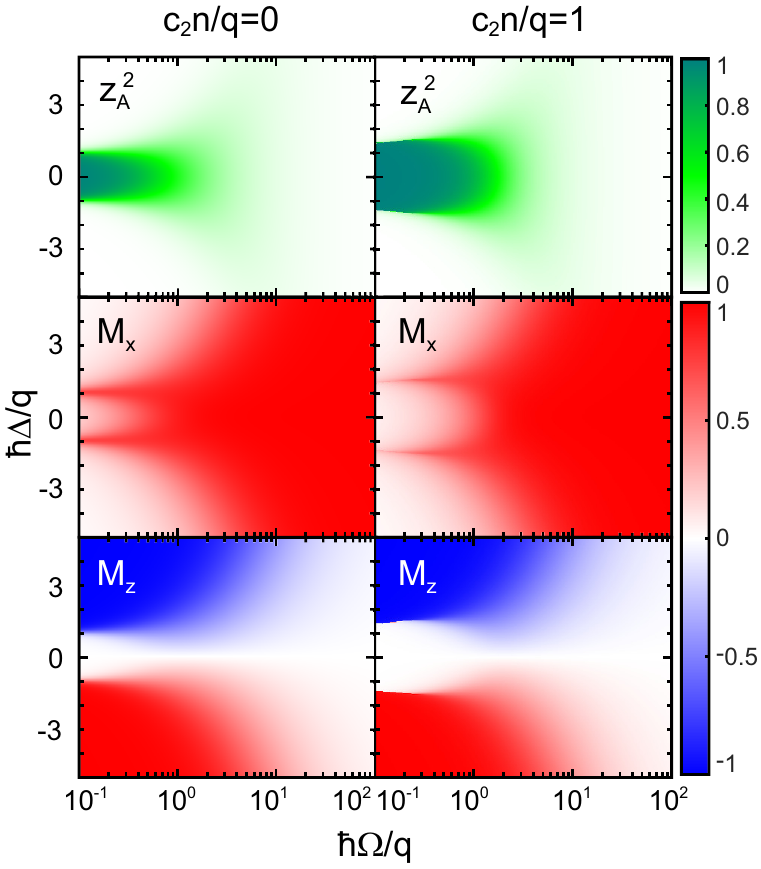}
    \caption{Phase diagram of the mean-field spin ground state of a homogeneous BEC under RF driving in the plane of $\Omega$ and $\Delta(=\omega-\omega_0)$ for $c_2n$=0 and $q$. $z_{A}^2$ is the polar fraction of the ground state~\cite{Oh14} and $M_{x,z}$ are the normalized magnetizations along $\hat{\bm{x}}$ and $\hat{\bm{z}}$, respectively.}
    \label{PhaseDiagram}
\end{figure}

To characterize the field noise effect in RF driving, we also investigated the stability of the spin transfer efficiency of a resonant RF pulse. In an ideal situation without field noises, atoms in the $m_z$=0 state can be completely transferred to the other spin states by a $\pi/2$ RF pulse, i.e., with a pulse duration of $\frac{\pi}{2\Omega}$. However, such a deterministic transfer is hampered by field noises because the spin rotation axis is modified to be along $\Omega \hat{\bm{x}} + \delta \hat{\bm{z}}$ which varies with realization. In Fig.~\ref{ACnoise}(c), we display the measurement results of the remaining population fraction of the $m_z$=0 state after a $\pi/2$ pulse for various $\Omega$. When $\Omega$ decreases to below $\delta_w$, the transfer efficiency decreases and fluctuates significantly. We numerically confirm that the measurement results are consistent with the field noise magnitude deduced from the spin-echo measurements.

\section{Mean-field phase diagram}\label{sec:Appendix2}

Under RF driving, the mean-field spin energy of the BEC is given by 
\begin{equation}
\label{Eq:DrivenE}
    E'_{\text{s}} = \Big[ \hbar \Delta \langle F_z \rangle  +q \langle F_z^2 \rangle -\hbar \Omega \langle F_x \rangle +\frac{c_2 n}{2} \langle \mathbf{F} \rangle^2 \Big] n,
\end{equation} 
where $\Delta$=$\omega$$-$$\omega_0$ is the frequency detuning of the RF field.
Based on this energy functional, we numerically calculated the stationary spin ground state of the system for a fixed $n$~\cite{Antoine14}. In Fig.~\ref{PhaseDiagram}, the phase diagrams for $c_2n =0$ and $q$ are presented in the plane of $\Omega$ and $\Delta$, where $z_{A}^2=\sqrt{1-|\langle \mathbf{F}\rangle|^2}$ is the polar fraction of the BEC~\cite{Oh14} and $M_{x,z}=\langle F_{x,z}\rangle$ are the normalized magnetizations along $x$ and $z$ directions, respectively. When the quadratic Zeeman energy is dominant, i.e., $ q/\hbar \gg\Delta$ and $\Omega $, the ground state is the EAP state with $\hat{\bm{d}}\parallel \hat{\bm{z}}$, whereas when the linear Zeeman energy is increased for $\Delta > q/\hbar$ ($\Omega >q/\hbar$), the system evolves to a ferromagnetic state polarized along $\hat{\bm{z}}$ ($\hat{\bm{x}}$). The parameter region of the polar state is extended with $c_2 n \neq 0$ because it is energetically more favored with the antiferromagnetic spin interactions.

\end{document}